\documentclass[hyper]{JHEP}
\usepackage{graphicx,amssymb,bm,latexsym,amsmath}
\usepackage{epstopdf}
\usepackage{caption}
\usepackage{subcaption}
\usepackage[toc,page]{appendix}
\newcommand{\bej}[1]{ \begin{equation}\label{#1} }
\newcommand{\eej}{\end{equation}}
\newcommand{\beaj}[1]{\begin{eqnarray}\label{#1} }
\newcommand{\eeaj}{\end{eqnarray}}
\newcommand{\eq}[1]{(\ref{#1})}
\def\ZZZ{{\hskip-3pt\hbox{ Z\kern-1.6mm Z}}}
\def\zzz{{\hskip-3pt\hbox{ z\kern-1mm z}}}

\newcommand{\bd}{\bar{\rm D}}

\newcommand{\N}{\frac{m_{2}}{k_{2}}-\frac{m_{1}}{k_{1}}}

\newcommand{\be}{\begin{equation}}
\newcommand{\ee}{\end{equation}}
\newcommand{\ben}{\begin{eqnarray}\displaystyle}
\newcommand{\een}{\end{eqnarray}}

\def\one{{\hbox{ 1\kern-.8mm l}}}
\def\zero{{\hbox{ 0\kern-1.5mm 0}}}

\def\be{\begin{equation}}       
\def\ee{\end{equation}}         

\def\bea{\begin{eqnarray}}      
\def\eea{\end{eqnarray}}
                                
\def\ba{\begin{array}}
\def\ea{\end{array}}
\def\bd{\begin{displaymath}}
\def\ed{\end{displaymath}}

\def\eq{\begin{equation}}
\def\eqe{\end{equation}}
\def\eqa{\begin{eqnarray}}
\def\eqae{\end{eqnarray}}
\def\ena{\end{eqnarray}}

\def\unit{1 \hskip-.3em \raise2pt\hbox{$ \scriptstyle |$ } }

\def\bd{\begin{displaymath}}
\def\ed{\end{displaymath}}

\def\6{\partial}
\def\N4{{\cal N}=4}

\def\bop#1{\setbox0=\hbox{$#1M$}\mkern1.5mu
        \vbox{\hrule height0pt depth.04\ht0
        \hbox{\vrule width.04\ht0 height.9\ht0 \kern.9\ht0
        \vrule width.04\ht0}\hrule height.04\ht0}\mkern1.5mu}

\def\>{\rangle} 

\def\<{\langle} 
\def\Dsl{D \hskip-.6em \raise1pt\hbox{$ / $ } }
\def\to{\rightarrow}

\def\+{\oplus}

\def\as2{AdS_3\times S^3_1 \times S^3_2}

\title{On multi-spin classical strings with NS-NS flux }
\author{Aritra Banerjee \\
CAS Key Laboratory of Theoretical Physics,~Institute of Theoretical Physics,\\ Chinese Academy of Sciences, Beijing 100 190, China

Email: \email{aritra@itp.ac.cn}}

\author{Sagar Biswas\\
Department of Physics, Ramakrishna Mission Vidyamandira, \\ Belur Math, Howrah 711 202, India

Email: \email{biswas.sagar@vidyamandira.ac.in}}

\author{Kamal L Panigrahi \\
 Department of Physics, Indian Institute of Technology Kharagpur,\\  Kharagpur 721 302, India

Email: \email{panigrahi@phy.iitkgp.ac.in}}
\abstract{We study multi spin semiclassical strings in $AdS_3\times S^3 \times T^4$ background supported by a mixture of Neveu-Schwarz-Neveu-Schwarz (NS-NS) and Ramond-Ramond (R-R) fluxes. This `mixed flux' background has been recently proved to be classically integrable. We start with a particular rigidly spinning fundamental string in $AdS_3\times S^1$ coupled to the NS-NS flux and classify the possible profiles. We also find out how the scaling relation among the energy and angular momenta of such a string changes due to presence of these fluxes. We emphasize on pure NS-NS flux case and discuss the fate of such solutions in that limit. }


\begin{document}
\section{Introduction}
In the two decades of its existence, the $AdS/CFT$ correspondence \cite{a} has turned out to be one of the most powerful tools of modern theoretical physics. Although the most successful version of the correspondence relies on the duality between type IIB string states in $AdS_5\times S^5$ and certain operators in a four dimensional Conformal Field Theory (CFT) living on the boundary of $AdS_5$, lower dimensional versions have also been well explored in the literature. Superstring theory on $AdS_3 \times S^3\times \mathcal{M}^4$ backgrounds have recently been studied extensively due to the renewed interest in $AdS_3/CFT_2$ correspondence \cite{b}\footnote{For a recent review of this vast subject, we redirect the reader to \cite{Sfondrini} and references therein.}. The most tractable cases of this duality with the compact manifold $\mathcal{M}^4$ being $T^4$ \cite{VIII,SST,AB,SW,Ads3S3T4A,Borsato:2013hoa,Ads3S3T4B,BianchiHoare,Spinconnection,%
BothM4,completeworldsheet,Ads3S3T4C,BMN mismatch,Ads3S3T4D,BPS} or $S^3\times S^1$ \cite{SS,Rughoonauth,SU11,BOS,Abbott,EGGL} have played a crucial role in understanding the duality. Although the dual $CFT_2$ has not been exactly found, there has been a lot of proposals \cite{LarsenMartinec, XIV, Tong, EberhardtGopakumar} which has helped to gain deep insight into the issue, leading to an exciting search in its own right. Another important property is that both of these above cases have been proven to be classically integrable \cite{BSZ,wulff}, which has called for direct integrability related approaches to study the duality parallel to the $AdS_5/CFT_4$ case  \cite{review}.

A very prominent difference from the $AdS_5\times S^5$ case, where only Ramond-Ramond (RR) five form fluxes are present, here both $AdS_3\times S^3\times T^4$ and $AdS_3\times S^3\times S^3\times S^1$   can be supported by both NS-NS and R-R three-form fluxes. For the former case, the integrability properties are not spoiled by these mixture of three-form fluxes \cite{CagnazzoZarembo} and the background indeed satisfies type IIB supergravity field equations, provided the parameters associated to field strength of NS-NS fluxes (say $b$) and the same for R-R fluxes (say $\hat b$) are related via $b^2 + \hat b^2 = 1$.  The  $AdS_3\times S^3\times T^4$ with `mixed' three-form fluxes has been an interesting  testing laboratory for the extension of $AdS_3/CFT_2$ duality for last few years \cite{BianchiHoare,HT,B.Hoare,Lloyd,SW2,M.Baggio,Modulispace}.  This background is conjectured to be originated from the near horizon geometry of the intersecting F1-NS5-D1-D5 branes in supergravity, although an explicit construction is still to be found. In the near horizon geometry the associated dilaton happens to be a constant. 

For the pure NS-NS flux which correspond to  $b = 1$, the  worldsheet theory can be understood by the $SL(2,\mathbb{R})$ Wess-Zumino-Witten (WZW) model \cite{Moo}.  Then the mixed flux model, must interpolate between a pure  R-R  flux case ($b=0$),  for  which  one  can  find  the  spectrum  using  the  integrability based approaches, and the pure NS-NS case ($b=1$) where the WZW approach has been quite successful \footnote{More recently, the spectrum and the S-matrix of the pure NS-NS WZW model has been discussed in detail \cite{purens1}. It has been reiterated that the pure WZW point corresponds to a integrable theory with linear dispersion relations. Moreover in \cite{purens2}, a spin-chain interpretation of the theory has been given. }.  But it has been clear that for a intermediate value of $b$, none of the approaches will be
suitable enough.  Studying classical strings and their dynamics in this background with mixed flux promises be helpful to relate these two extreme models together. The mixed theory of course has an elegant formulation in terms of the supercoset construction and large amount of work has been devoted to the study of classical integrability, and the S-matrix construction to name a few. Though a compete understanding of the dynamics in the form of dual theory is still lacking, it seems an interesting background to work with especially to understand various closed and open fundamental string solutions which in turn can teach us about the relevant dual operators. 

Over the last few years, the integrability of quantum string theory in $AdS_5\times S^5$ and its various cousins has led to remarkable progress in our understanding of the spectrum of anomalous dimensions in the dual known gauge theories. In this connection, the study of string spectrum on semisymmetric superspaces have benefitted greatly from our understanding of classical integrability. Most notably, techniques related to semiclassical quantization \cite{FirstQuantizing, Quantizing} of spinning strings have played  an extremely important role, since the dynamics in this limit is invariably defined by well-known integrable systems. The $AdS_3 \times S^3 \times T^4$ background with mixed flux was shown to be classically integrable as we have mentioned before. Once the integrability of the background was proved, a modified `Giant Magnon' dispersion relation was proposed in \cite{B.Hoare}, which turned out to have non-periodic contribution in the momentum. Subsequently finite-gap equations were proposed \cite{FinitegapBabichenko,masslessfinitegap} and a large class of folded \cite{Rotating1}, rotating \cite{both, C.Ahn}, pulsating \cite{Rotating2, Barik1} and GKP-like multi-spike \cite{multispike} strings were studied in an attempt to formalise the AdS/CFT dictionary to have a hint of the relevant operators. In fact the presence of the fluxes seem to have modified scaling relations between conserved charges considerably to the leading order via terms depending on the parameter $b$. In another avenue of exploration, it was found that inclusion of the mixed flux adds an integrable deformation to the well known Neumann-Rosochatius dynamical model of strings in $AdS_3\times S^3$ and relevant classical solutions \cite{HN1,HN2,HN3} and one-loop quantization \cite{HN4} has been elucidated in the literature. Moreover, probe D1 strings exploring the mixed flux background have been discussed in \cite{D1}.

Motivated by the interest in understanding more details of the string geodesics in the presence of flux, in this paper we make a modest attempt of studying a classical rigidly rotating string  in  $AdS_3\times S^1$ background with mixed fluxes. This geometry allows us to study effect of an extra angular momentum along $S^1$. Previously, in the context of analyzing the fundamental string in pure $AdS_3 \times S^1$ background, the ``Giant Magnons'' in the $SL(2)$ sector has been well studied \cite{Minahan1}. In the dual spin chain picture, these magnons are also associated to `long' gauge invariant operators. In fact it was speculated that these strings correspond to some Wilson lines
in particular representation along a trajectory of the gauge theory. 

The rest of the paper is organised as follows. In section 2, we study a general rigidly rotating strings in $AdS^3\times S^1 \subset AdS_3\times S^3$ with both kinds of fluxes present. We completely classify possible configurations of the string and find the associated profiles integrating the equations of motion. In section 3, we write down the conserved charges for the string motion, and describe how the diverging charges can be regulated. Then, using different consistent values of our parameters, we arrive at regularised dispersion relations for these charges, which can be directly compared with their $\mathbb{R}\times S^3$ counterparts. To be complete, we have separately discussed the case of pure NS-NS flux in each of the above sections.
 In section 4, we comment on the general `soliton' structure of these solutions. In section 5 we summarise and conclude our work. 
\section{Rotating strings in $AdS_3\times S^1$ in the presence of B field}
Consider the $AdS_3 \times S^1$ metric and the two-form NS-NS and R-R fields given by,
\begin{eqnarray}
    ds^2 &=& -\cosh^2\rho ~dt^2 + d\rho^2 + \sinh^2\rho~ d\phi^2 + d\varphi^2, \nonumber \\
     B_{t\phi} &=& b\sinh^2\rho~ dt \wedge d\phi \ ,~~ C_{t\phi} = \sqrt{1-b^2}\sinh^2\rho~ dt \wedge d\phi  \nonumber \\ 
\end{eqnarray}

Since we are interested in a fundamental string, the R-R flux does not couple to it. We emphasise here that there is a gauge freedom in the choice of the two-form NS-NS-field since the conformal
invariance conditions or, to leading order, the supergravity equations of motion, instead depend on the three-form field strength, $H^{(3)}=d B$. For details of this ambiguity one can look at \cite{B.Hoare}, where, for example, the 2-form field in $\mathbb{R}\times S^3$ is written as 
$-\frac{b}{2}(\cos 2\theta + \chi)$, with $\chi$ is a constant that gives rises to a constant ambiguity term in the WZ part of the action. This constant $\chi$ can be chosen via imposing particular physical
requirements on the classical string itself. In our case this extra constant will only generate a constant shift in the desired  dispersion relations, and thus can be chosen suitably to give the expressions mentioned above. 

In this background, to study a fundamental string coupled to an antisymmetric NS-NS B-field, we use the Polyakov action,
\begin{eqnarray}
S=-\frac{\sqrt{\lambda}}{4\pi}\int d\sigma d\tau
[\sqrt{-\gamma}\gamma^{\alpha \beta}g_{MN}\partial_{\alpha} X^M
\partial_{\beta}X^N - \epsilon^{\alpha \beta}\partial_{\alpha} X^M
\partial_{\beta}X^N B_{MN}] \ ,
\end{eqnarray}
where $\lambda$ is the 't Hooft coupling,
$\gamma^{\alpha \beta}$ is the worldsheet metric and $\epsilon^{\alpha
	\beta}$ is the antisymmetric tensor defined as $\epsilon^{\tau
	\sigma}=-\epsilon^{\sigma \tau}=1$.
Variation of the action with respect to
$X^M$ gives us the following equations of motion
\begin{eqnarray}
2\partial_{\alpha}(\eta^{\alpha \beta} \partial_{\beta}X^Ng_{KN})
&-& \eta^{\alpha \beta} \partial_{\alpha} X^M \partial_{\beta}
X^N\partial_K g_{MN} - 2\partial_{\alpha}(\epsilon^{\alpha \beta}
\partial_{\beta}X^N b_{KN}) \nonumber \\ &+& \epsilon ^{\alpha \beta}
\partial_{\alpha} X^M \partial_{\beta} X^N\partial_K b_{MN}=0 \ ,
\end{eqnarray}
and variation with respect to the metric gives the two Virasoro
constraints,
\begin{eqnarray}
g_{MN}(\partial_{\tau}X^M \partial_{\tau}X^N +
\partial_{\sigma}X^M \partial_{\sigma}X^N)&=&0 \ ,\label{v1} \\ 
g_{MN}(\partial_{\tau}X^M \partial_{\sigma}X^N)&=&0 \label{v2}\ .
\end{eqnarray}
We use the conformal gauge (i.e.
$\sqrt{-\gamma}\gamma^{\alpha \beta}=\eta^{\alpha \beta}$) with
$\eta^{\tau \tau}=-1$, $\eta^{\sigma \sigma}=1$ and $\eta^{\tau
\sigma}=\eta^{\sigma \tau}=0$) to solve the equations of motion.
For studying a generic class of rotating strings we use the ansatz,
\begin{eqnarray}
    t = \tau + g_1(y) \ ,  \rho = \rho(y) \ ,  \phi = \omega(\tau + g_2(y)) \ ,  \varphi = \mu\tau \ .
\end{eqnarray}
where the generalised variable $y = \sigma - v\tau$, which takes worldsheet coordinates from $(\sigma,\tau)$ to $(y,\tau)$.
Solving $t$ and $\phi$ equations we get,
\begin{eqnarray}
    \frac{\partial g_1}{\partial y} = \frac{1}{(1-v^2)}\Big[ \frac{b\omega\sinh^2\rho - c_1}{\cosh^2\rho} - v\Big] \ ,  \frac{\partial g_2}{\partial y} = \frac{1}{(1-v^2)}\Big[ \frac{b\sinh^2\rho + \omega c_2}{\omega\sinh^2\rho} - v\Big] \ .
\end{eqnarray}
$c_{1,2}$ are constants here. And solving for $\rho$ equation we get,
\begin{equation}\label{chieq}
    \Big(\frac{\partial \rho}{\partial y}\Big)^2 = \frac{1}{1-v^2}\Big[(1-b^2)(1-\omega^2)\cosh^2\rho + \frac{(c_1 + b\omega)^2}{\cosh^2\rho} - \frac{\omega^2 c_2^2}{\sinh^2\rho}\Big] + c_3 \ .
\end{equation}
where $c_1$, $c_2$ and $c_3$ are again the integration constants.
Comparing the two Virasoro constraints, we get the following relation between various constants,
\begin{equation}
    c_1 + c_2\omega^2 + \mu^2v = 0 \ .
\end{equation}
Using the boundary condition in the limit, $\frac{\partial\rho}{\partial y} \to 0$ as $\rho \to 0$ implies $c_2=0$ and also
\be
c_3 = -\frac{1}{(1-v^2)}[(1-b^2)(1-\omega^2) + (c_1 + b\omega)^2]
\ee
 Putting in the conditions, we get from (\ref{chieq}),
\begin{equation}\label{profile}
    \frac{\partial\rho}{\partial y} = \pm \frac{\sqrt{(1-b^2)(1 - \omega^2)}}{(1 - v^2)} \tanh\rho \sqrt{\cosh^2\rho - \frac{(c_1 + b\omega)^2}{(1-b^2)(1-\omega^2)}} \ .
    \end{equation}
    
We now move on to discuss the various classes of strings that occur here for different values of the parameters introduced. 
  Using the above values of integration constants we get the relation,
\begin{equation}
    c_1^2 + \frac{1+v^2}{v}c_1 + 1 = 0.
\end{equation}
Solving this equation we get $c_1 = -v$ or $c_1 = -\frac{1}{v}$.
For $c_1 = -\frac{1}{v}$ implies $\mu = \frac{1}{v}$ again for $c_1 = -v$ implies $\mu = 1$ via the relation $c_1+\mu^2 v = 0$.
  
  Let's start with the case of $c_1 = -v$, where we can demand forward propagation of these strings,
  \begin{equation}
    \dot{t} = \frac{1}{1-v^2}\Big[1 - b\omega v + \frac{v(b\omega -v)}{\cosh^2\rho}\Big] \ge 0.
\end{equation}
which implies $\rho > \cosh^{-1}(\sqrt{\frac{v(b\omega - v)}{1 - b\omega v}})$ . Now, solving for the string profile equation we get the following expression,
\begin{equation}
    y = \pm \frac{(1 - v^2)}{\sqrt{(1-b^2)(1-\omega^2)(1-\alpha^2)}} \tanh^{-1}\Big(\sqrt{\frac{\cosh^2\rho - \alpha^2}{1-\alpha^2}}\Big)
\end{equation}
where $\alpha = \frac{b\omega - v}{\sqrt{(1-b^2)(1-\omega^2)}}$ is one root of the equation $\rho'(y)= 0$. Depending on the value of this root (let's call it $\cosh^2\rho_1$), we could have different profiles.\\
$~~~~~~~~$
\textbf I) When $\cosh^2\rho_1>1$, the numerator inside $(1-\alpha^2)$ part is negative, and we could write the profile equation in the trigonometric form
\be
y = \pm\frac{1}{\beta}\cos^{-1}\left( \frac{\sqrt{b^2+\omega^2+v^2-2b\omega v-1}}{\sqrt{(1-b^2)(1-\omega^2)}\sinh\rho}   \right)
\ee
Evidently, these profiles are valid between $-\frac{\pi}{2\beta}\leq y \leq \frac{\pi}{2\beta}$, where 
\be
\beta= \frac{\sqrt{b^2+\omega^2+v^2-2b\omega v-1}}{1-v^2} .
\ee
Usually these string profiles correspond to the so-called `Hanging' strings. It is easy to see at the boundaries $y =\pm\frac{\pi}{2\beta}$, the radial coordinate $\rho$ goes to infinity, i.e. attaches to the $AdS$ `boundary', and at $y=0$ it takes the minimal value, earning this class of profiles the name as we mentioned.
 \begin{figure}[!h]
		\centering
		\includegraphics[width=10.5cm]{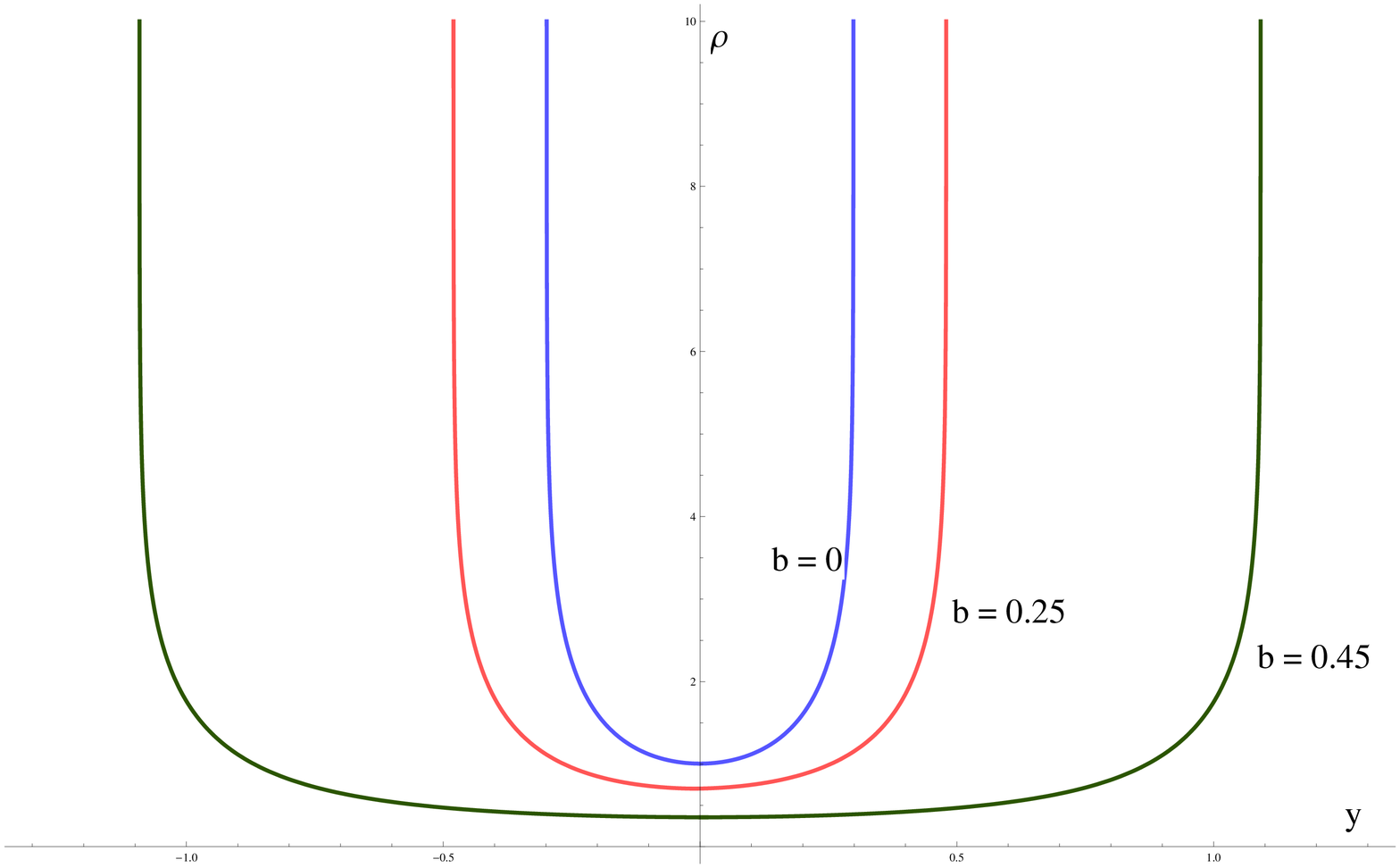}
		\caption{Plot of hanging string profiles with $\omega = 0.8$ and $v=0.93$, notice that with increasing values of $b$ the area enclosed by the string increases rapidly.}
\end{figure}
\bigskip
\\$~~~~$\textbf{II}) The other class of solutions are obtained when we use $\cosh^2\rho_1\leq1$, for which the profile can be written in the following form,
\be
y = \pm \frac{1-v^2}{\sqrt{1-b^2-\omega^2-v^2+2bv\omega}}\sinh^{-1}\left(\frac{\sqrt{1-b^2-\omega^2-v^2+2bv\omega}}{\sqrt{(1-b^2)(1-\omega^2)}\sinh\rho}  \right).
\ee
As we can see, the range of $y$ here changes from the case before to $-\infty\leq y \leq \infty$, and these profiles are particularly called `Spiky' strings. These solutions are highly interesting due to this infinite range of $y$, and presence of special points with derivative discontinuities on the worldsheet. Take note that the above solution can be written in the following form,
\bea
\sinh\rho&=& \frac{\hat\gamma}{\sinh\hat\beta y},~~~~~~~~0\leq y\leq \infty \nonumber \\
&=&  -\frac{\hat\gamma}{\sinh\hat\beta y},~~~-\infty\leq y\leq 0. \nonumber \\
\eea
Where the parameters are given by,
\be
\hat\gamma = \frac{\sqrt{1-b^2-\omega^2-v^2+2bv\omega}}{\sqrt{(1-b^2)(1-\omega^2)}} ,~~~\hat\beta=  \frac{\sqrt{1-b^2-\omega^2-v^2+2bv\omega}}{1-v^2}.
\ee
One might notice here that $\hat\gamma =\sqrt{1-\alpha^2}$ and $\beta = i\hat\beta$. We will be using the above solution repeatedly throughout this work to study various cases.
We can notice from the figures 1 and 2, that as we increase the value of the NS-NS flux, i.e. take $b$ towards $1$, the strings tend to become wider. This `fattening' of the strings under influence of NS-NS flux has been noticed before in \cite{multispike}.  
\begin{figure}[!htb]
\centering
\begin{minipage}{.5\textwidth}
 \subsection*{}
  \centering
  \includegraphics[width=7.5cm, height=7cm]{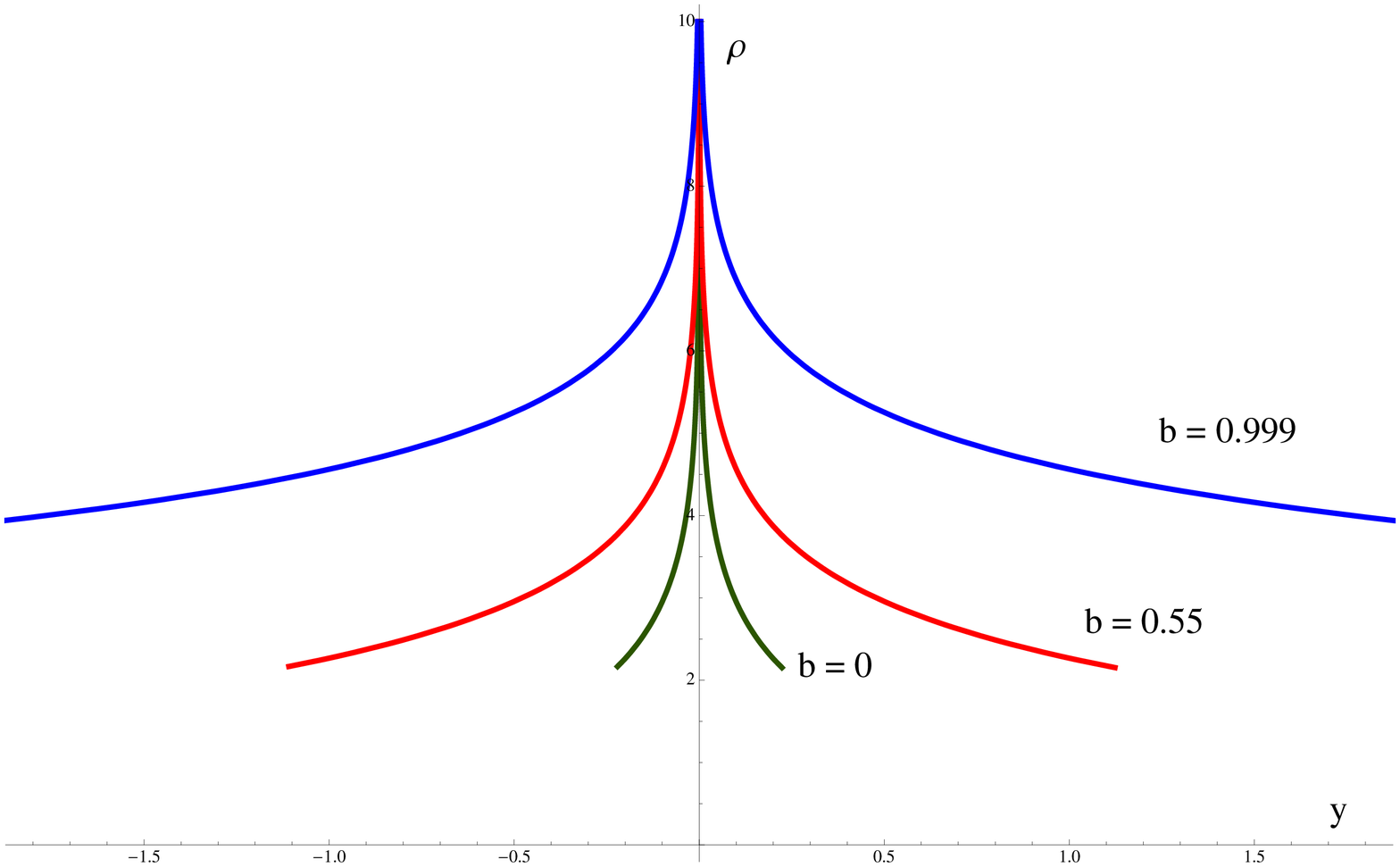}
  \subcaption{}
  \label{fig:test1}
\end{minipage}
\begin{minipage}{.5\textwidth}
  \centering
  \includegraphics[width=1.02\linewidth]{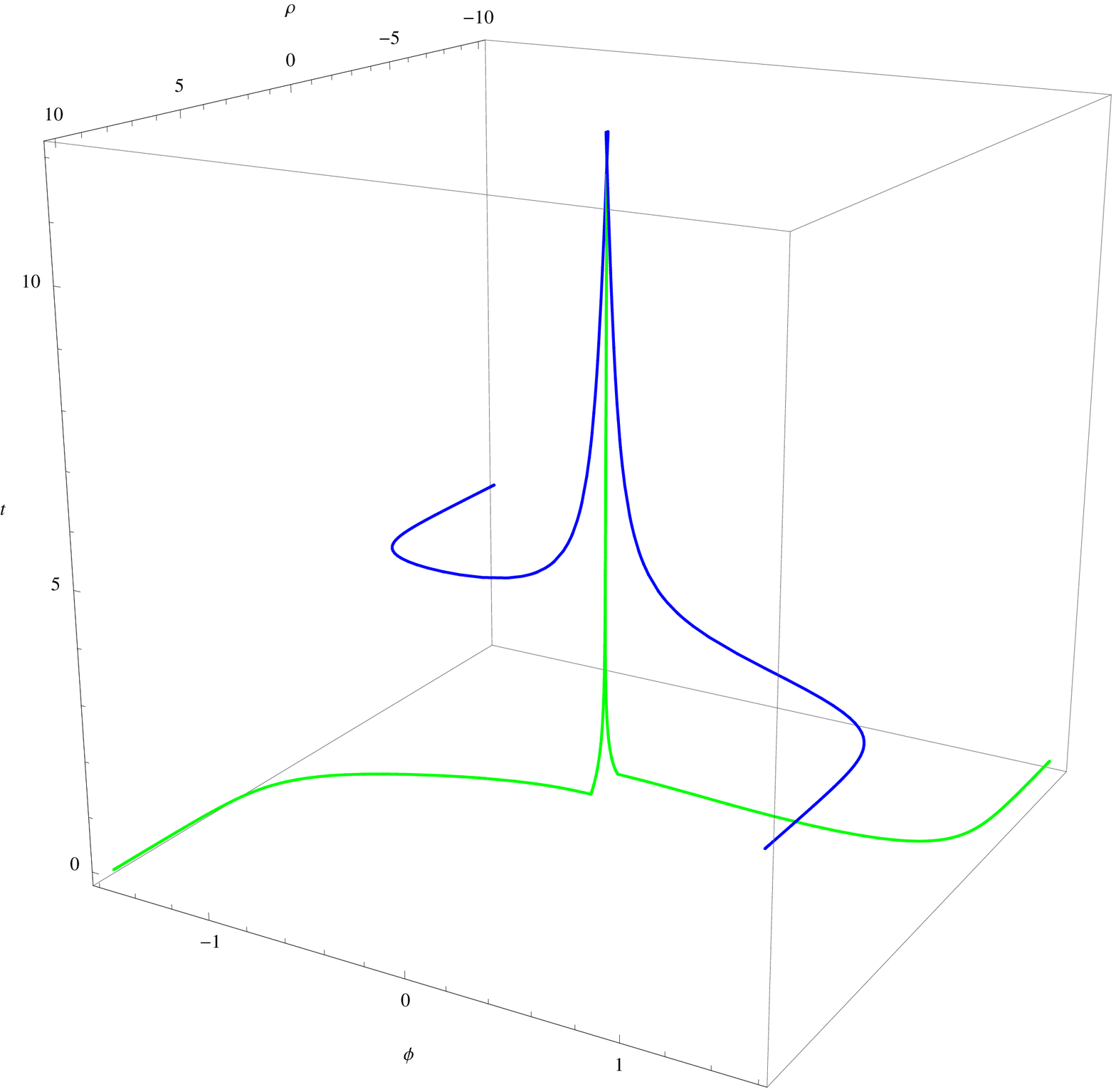}
  \subcaption{}
  \label{fig:test2}
\end{minipage}
\caption{Plot of `Spiky' strings, (a) Strings in the $\rho-y$ plane with $\omega=0.55$ and $v=0.45$. Notice that with increasing $b$, the spikes become wider, which we refer to as `fattening' of the spike. (b) A $3d$ view of the fattening of spike in $(t,\rho,\phi)$ space for $b=0$ (green) and $b=0.2$ (blue). The snapshot is taken at $\tau= 0$.  }
\end{figure}
\bigskip
\\Let us, for completeness also discuss the case of $c_1 = -\frac{1}{v}$, although we will be mostly interested in the previous case as mentioned above.  In this case we again could demand forward propagation of strings, i.e.
\begin{equation}
    \dot{t} =\frac{1 - v^2-b\omega v}{1-v^2}\tanh^2\rho \ge 0 \ ,
\end{equation}
which implies $1 -v^2- b\omega  v> 0$ and solving for the string profile from the $\rho$ equation we get,
\begin{equation}
    y = \pm \frac{(1 - v^2)}{\sqrt{(1-b^2)(1-\omega^2)(1-\alpha_1^2)}} \tanh^{-1}\Big(\sqrt{\frac{\cosh^2\chi - \alpha_1^2}{1-\alpha_1^2}}\Big)
\end{equation}
where we have $\alpha_1 = \frac{ b\omega v-1}{v\sqrt{(1-b^2)(1-\omega^2)}}$, where we would call the constant $\alpha_1^2 = \cosh^2\rho_2$. Now to look at the parameter space, we see while the $b=0$, (i.e pure RR case)  these strings will only give rise to one class of strings since $\alpha_1^2[b=0]\geq 1$, in our case it is not true. So, as we discussed in the earlier case, there could be both `hanging' and `spiky' strings present for these values of the parameters. 
\subsection*{The case of $b=1$: Pure NS-NS flux}
As we discussed earlier, pure NS-NS case is important in the sense the strings in this case are described by a $SL(2,\mathbb{R})$ WZW model. Pure NS-NS case can be achieved by putting $b=1$, but just putting that carelessly into the expressions we derived in the last section would simply lead to diverging solutions. So we start with the $t$ and $\phi$ equations in the $b=1$ case, which give,
\begin{eqnarray}
    \frac{\partial g_1}{\partial y} = \frac{1}{(1-v^2)}\Big[ \frac{\omega\sinh^2\rho - c_1}{\cosh^2\rho} - v\Big] \ ,  \frac{\partial g_2}{\partial y} = \frac{1}{(1-v^2)}\Big[ \frac{\sinh^2\rho + \omega c_2}{\omega\sinh^2\rho} - v\Big] \ ,
\end{eqnarray}
where $c_{1,2}$ are constants as before. Solving for $\rho$ equation of motion we get,
\begin{equation}
    \Big(\frac{\partial \rho}{\partial y}\Big)^2 = \frac{1}{(1-v^2)^2}\Big[ \frac{(c_1 + \omega)^2}{\cosh^2\rho} - \frac{\omega^2 c_2^2}{\sinh^2\rho}\Big] + c_3 \ .
\end{equation}
where $c_1$, $c_2$ and $c_3$ are the integration constants.
Comparing the two Virasoro constraints, we get the following relation between various constraints,
\begin{equation}
    c_1 + c_2\omega^2 + \mu^2v = 0 \ .
\end{equation}
and comparing with the Virasoro we get,
\be \label{const}
c_3 = \frac{1}{(1-v^2)^2} [1 - \omega^2 - 2\omega(c_1 + c_2) + (c_1+ c_2\omega^2)\frac{1+v^2}{v}] \ .
\ee
In this case we can't use the boundary condition that $\frac{\partial\rho}{\partial y} \to 0$ as $\rho \to 0$, as it leads to  imaginary solutions for the string profile, which would be seemingly unphysical. Instead, since we still want the $\frac{\partial\rho}{\partial y} $ to be regular at $\rho = 0$, we still demand $c_2= 0$, which, in conjunction with \ref{const}, leads us to,
\be
\left(\frac{\partial\rho}{\partial y} \right)^2 =\frac{\big[ (c_1+v)(c_1+\frac{1}{v}) \big]}{(1-v^2)^2}\frac{(\cosh^2\rho - k^2 \sinh^2\rho)}{\cosh^2\rho},
\ee
where we have the constant,
\be
k^2 = \frac{(c_1+\omega)^2}{\big[ (c_1+v)(c_1+\frac{1}{v}) \big]} \ .
\ee
So, we could see that one can't put $c_1= -v$ or $-\frac{1}{v}$  as it would make $\frac{\partial\rho}{\partial y} $ imaginary.
Equivalently, by putting $c_1 = -\mu^2 v$ we obtain the form,
\be
\frac{\partial\rho}{\partial y}  = \pm \frac{\sqrt{(1 - \mu^2)(1 - \mu^2 v^2)} }{1-v^2}\frac{\sqrt{\cosh^2\rho - k^2 \sinh^2\rho}}{\cosh\rho},
\ee
where, now the constant can be written as
$
k = \frac{\omega - \mu^2v}{\sqrt{(1 - \mu^2)(1 - \mu^2v^2)}} \ .
$
Integrating the above expression, we can get the profile in the form,
\be
 y  = \pm \frac{1-v^2}{\sqrt{(1 -\mu^2-\mu^2 v^2 -\omega^2+2\mu^2 v\omega)}}\sinh^{-1}\left[\sqrt{1-k^2}~\sinh\rho    \right] \ .
\ee
This string profile clearly has a turning point at $k=1$. We can expand $k$ near the turning point as $k=1+2\eta,~~\eta<<1$, which leads us to a solution of the form,
\be
\rho(y) \sim \sinh^{-1}\Big[ \frac{\sin 2\sqrt{\eta} (y-y_0)}{2\sqrt{\eta}}\Big],
\ee
sans some constant factors. Since $\eta$ is small, this points to a slowly oscillating strings near the turning point with very large periods in $y$. 
\section{Conserved Charges and Regularised Dispersion Relations}
To determine the conserved charges associated to the string motion, we start from the full form of the sigma model action,
\begin{eqnarray}
   S &=& -\frac{\sqrt{\lambda}}{4\pi} \int d\tau d\sigma \Big[-\cosh^2\rho [(\partial_{\sigma}t)^2 - (\partial_{\tau}t)^2] + [(\partial_{\sigma}\rho)^2 - (\partial_{\tau}\rho)^2] + \sinh^2\rho [(\partial_{\sigma}\phi)^2 - (\partial_{\tau}\phi)^2] \nonumber \\ && + [(\partial_{\sigma}\varphi)^2 - (\partial_{\tau}\varphi)^2] + 2b\sinh^2\rho(\partial_{\sigma}t \partial_{\tau}\phi - \partial_{\tau}t\partial_{\sigma}\phi)\Big]
\end{eqnarray}
From this action we can easily determine the conserved energy and spins as following,
\begin{eqnarray}
    E &=& - \int \frac{\partial\mathcal{L}}{\partial\dot{t}} d\sigma = \frac{\sqrt{\lambda}}{2\pi} \int[\cosh^2\rho \dot{t} - b\sinh^2\rho \phi^{\prime}] ~d\sigma \nonumber \\ && = \frac{\sqrt{\lambda}}{2\pi (1 - v^2)} \int [\cosh^2\rho - b^2 \sinh^2\rho + c_1 v - c_2 b \omega]~ d\sigma \ , \nonumber \\
J_{\phi} &=&  \int \frac{\partial\mathcal{L}}{\partial\dot{\phi}} ~d\sigma = \frac{\sqrt{\lambda}}{2\pi} \int[\sinh^2\rho (\dot{\phi} - b t^{\prime})] ~d\sigma \nonumber \\ && = \frac{\sqrt{\lambda}}{2\pi (1 - v^2)} \int \Big[\omega\sinh^2\rho - c_2\omega v - \frac{b\sinh^2\rho}{\cosh^2}(b\omega\sinh^2\rho - c_1)\Big] ~d\sigma \ , \nonumber \\  J_{\varphi} &=&  \int \frac{\partial\mathcal{L}}{\partial\dot{\varphi}} ~d\sigma = \frac{\sqrt{\lambda}}{2\pi} \mu \int d\sigma \ , \nonumber \\
\end{eqnarray}
Here $J_\phi$ and $J_\varphi$ are spins associated to the angular directions $\phi$ and $\varphi$. And the time difference $\Delta t$, between the endpoints of the string is given by
\begin{equation}
    \Delta t = \int\frac{\partial g_1}{\partial y} d\sigma = \frac{1}{1-v^2} \int \Big[\frac{b\omega\sinh^2\rho - c_1}{\cosh^2\rho} - v\Big] d\sigma \ .
\end{equation}
In what follows, we will discuss about the fate of these charges for the two cases we discussed before, i.e. $c_1 = -v$ and $c_1 = -\frac{1}{v}$, where $c_2 = 0$. These two distinct cases in the parameter space will lead us to distinct dispersion relations between the charges.
\subsection{Case I:  $c_1 = -v$, $c_2 = 0$ and $\mu = 1$}
 In this case we find that (\ref{profile}) becomes,
\begin{equation}
    \frac{\partial \rho}{\partial y} = \frac{\sqrt{(1 - b^2)(1 - \omega^2)}}{1 - v^2} \tanh \rho \sqrt{\cosh^2\rho - \alpha^2} \ ,
\end{equation}
where we remind the reader that $\alpha = \frac{b\omega - v}{\sqrt{(1 - b^2)(1 - \omega^2)}}$.
The conserved charges in this case takes the form,
\begin{eqnarray}
    E = \frac{\sqrt{\lambda}}{\pi\sqrt{(1-b^2)(1-\omega^2)}} && \Big[(1-b^2) \int_0^{\infty} \frac{\sinh\rho \cosh\rho d\rho}{\sqrt{\cosh^2\rho - \alpha^2}} \nonumber \\ && + (1-v^2)\int_0^{\infty} \frac{\cosh\rho d\rho}{\sinh\rho \sqrt{\cosh^2\rho - \alpha^2}} \Big] \ , \nonumber \\ J_{\phi} = \frac{\sqrt{\lambda}}{\pi\sqrt{(1-b^2)(1-\omega^2)}} && \Big[\omega(1-b^2) \int_0^{\infty} \frac{\sinh\rho \cosh\rho d\rho}{\sqrt{\cosh^2\rho - \alpha^2}} \nonumber \\ && + b(b\omega - v)\int_0^{\infty} \frac{\sinh\rho d\rho}{\cosh\rho \sqrt{\cosh^2\rho - \alpha^2}} \Big] \ , \nonumber \\ J_{\varphi} = \frac{\sqrt{\lambda} (1 - v^2)}{\pi\sqrt{(1-b^2)(1-\omega^2)}} && \int_0^{\infty} \frac{\cosh\rho d\rho}{\sinh\rho\sqrt{\cosh^2\rho - \alpha^2}} \ . \nonumber \\
\end{eqnarray}
Thus, we can easily see that all of the charges above diverge. Note here  $E$ contains both UV and IR divergences, $J_\phi$ has only UV divergence, while $J_{\varphi}$ contains only IR divergence \footnote {The integrals in $E$ with $\cosh\rho =r$ has forms 
$I_1 = \sqrt{r^2 - \alpha^2}~|_{1}^{\infty},~~~I_2 = \frac{1}{\sqrt{\alpha^2-1}}\tan^{-1}\Big[  \sqrt{\frac{r^2-\alpha^2}{\alpha^2-1}} \Big]_{1}^{\infty}$}
  The time difference between two end points of the string is however, finite, and is given by,
\begin{equation}
    \Delta t = \frac{(b\omega - v)}{\sqrt{(1 - b^2)(1 - \omega^2)}} \int_{-\infty}^{\infty} \frac{\sinh\rho d\rho}{\cosh\rho\sqrt{\cosh^2\rho - \alpha^2}} = 2\Big[\frac{\pi}{2} - \tan^{-1} \frac{\sqrt{1 - \alpha^2}}{\alpha} \Big] \ ,
\end{equation}
which implies $\alpha = \sin(\frac{\Delta t}{2})$. 
Now it is evident that $E - J_{\varphi}$ has the IR divergences are mutually cancelled, while the UV divergent part in $E$ remains, which we can take care of by introducing an UV cutoff $\Lambda$,
\begin{equation}
    E - J_{\varphi} = \frac{\sqrt{\lambda}\sqrt{1 - b^2}}{\pi\sqrt{1 - \omega^2}} \Big[ \Lambda-\sqrt{1 - \alpha^2} \Big] \ . \nonumber \\ 
\end{equation}
We can subtract away the extra divergent piece containing $\Lambda$ to write down a regularised version of this. Now we turn to the charge $J_\phi$, where curiously one must note that in $b=0$ (pure RR) case, the extra divergent piece does not occur. And hence, in the finite NS-NS flux case, one must consider a finite regularised combination having the following form,
\be\label{bigj}
\mathcal{J}_\phi=\Big(J_{\phi} - \frac{ \sqrt{\lambda}}{2\pi}b \Delta t \Big)_{reg} = -\frac{\omega\sqrt{\lambda}\sqrt{1 - b^2}}{\pi\sqrt{1 - \omega^2}} \sqrt{1 - \alpha^2} \ .
\ee
Now these conserved charges hold the relationship 
\begin{equation}
   ( E - J_{\varphi})_{reg} = \frac{1}{\omega}\mathcal{J}_\phi \ .
\end{equation}
We can replace this $\omega$ from (\ref{bigj}) and find these charges to satisfy the following dispersion relation,
\begin{equation}\label{magnon}
    (E - J_{\varphi})_{reg} = \sqrt{(J_{\phi} -  \frac{\sqrt{\lambda}}{2\pi}b \Delta t)_{reg}^2 + \frac{\lambda(1 - b^2)}{\pi^2} \cos^2\frac{\Delta t}{2}} \ .
\end{equation}
For $b = 0$ the above relation reduces to the well known result in the literature \cite{Minahan1}.  The above dispersion relation can be compared with the Giant Magnon solution in $\mathbb{R}\times S^3$ with NS-NS fluxes discussed in \cite{B.Hoare}. This relation has the same structure and can be interpreted as a bound state of $\mathcal{J}_\phi$ number of magnons moving with a momentum $\Delta t$ on the $AdS$ hyperboloid. The relation reduces to the one discussed in for the case of $b=0$. However there is no apparent periodicity in $\Delta t$ in this case.
\subsection{Case II: $c_1 = -\frac{1}{v}$, $c_2 = 0$ and $\mu = \frac{1}{v}$}
We discuss here the other case for completeness and providing clarity. In this case we would have from the profile equation,
\begin{equation}
    \frac{\partial \rho}{\partial y} = \frac{\sqrt{(1 - b^2)(1 - \omega^2)}}{1 - v^2} \tanh \rho \sqrt{\cosh^2\rho - \alpha_1^2} \ ,
\end{equation}
where $\alpha_1 = \frac{bv\omega - 1}{v\sqrt{(1 - b^2)(1 - \omega^2)}}$.
The expressions for conserved charges read,
\begin{eqnarray}
    E = \frac{\sqrt{\lambda}\sqrt{(1-b^2)}}{\pi\sqrt{(1-\omega^2)}} && \int_0^{\infty} \frac{\sinh\rho \cosh\rho d\rho}{\sqrt{\cosh^2\rho - \alpha_1^2}}  \ , \nonumber \\ J_{\phi} = \frac{\sqrt{\lambda}}{\pi\sqrt{(1-b^2)(1-\omega^2)}} && \Big[\omega(1-b^2) \int_0^{\infty} \frac{\sinh\rho \cosh\rho d\rho}{\sqrt{\cosh^2\rho - \alpha_1^2}} \nonumber \\ && - \frac{b(1 - b\omega v)}{v} \int_0^{\infty} \frac{\sinh\rho d\rho}{\cosh\rho \sqrt{\cosh^2\rho - \alpha_1^2}} \Big] \ , \nonumber \\ J_{\varphi} = \frac{\sqrt{\lambda} (1 - v^2)}{\pi v\sqrt{(1-b^2)(1-\omega^2)}} && \int_0^{\infty} \frac{\cosh\rho d\rho}{\sinh\rho\sqrt{\cosh^2\rho - \alpha_1^2}} \ . \nonumber \\
\end{eqnarray}
In this case $E$ has only UV divergence, $J_{\varphi}$ has only IR divergence. However, the time difference is also quite complicated here and has the following expression,
\begin{eqnarray}
    \Delta t = \frac{1}{v\sqrt{(1-b^2)(1-\omega^2)}} && \Big[(1 - v^2) \int_{-\infty}^{\infty} \frac{\cosh\rho d\rho}{\sinh\rho \sqrt{\cosh^2\rho - \alpha_1^2}} \nonumber \\ && - (1 - b \omega v) \int_{-\infty}^{\infty} \frac{\sinh\rho d\rho}{\cosh\rho \sqrt{\cosh^2\rho - \alpha_1^2}} \Big] \ .
\end{eqnarray}
This clearly has divergences, although we can find regularised time  difference by considering a combination of the charges, 
\begin{equation}
    (\Delta t)_{reg} = \Delta t - \frac{2\pi}{\sqrt{\lambda}} J_{\varphi} = 2\Big[\frac{\pi}{2} - \tan^{-1} \Big(\frac{\sqrt{1 - \alpha_1^2}}{\alpha_1}\Big) \Big] \ ,
\end{equation}
which implies $\alpha_1 = \sin(\frac{(\Delta t)_{reg}}{2})$.
Following the previous section, we can again find the following finite regularised quantities,
\begin{eqnarray}
    E_{reg}  &=& -\frac{\sqrt{\lambda}\sqrt{1 - b^2}}{\pi\sqrt{1 - \omega^2}} \sqrt{1 - \alpha_1^2} \ , \nonumber \\ J_{\phi} - \frac{b \sqrt{\lambda}}{2\pi} (\Delta t)_{reg} &=& -\frac{\omega\sqrt{\lambda}\sqrt{1 - b^2}}{\pi\sqrt{1 - \omega^2}} \sqrt{1 - \alpha_1^2} \ . \nonumber \\
\end{eqnarray}
These conserved charges satisfy the relation,
\begin{equation}
    E_{reg} = \frac{1}{\omega} \Big[ J_{\phi} - \frac{\sqrt{\lambda}}{2\pi} (b\Delta t)_{reg}\Big] \ ,
\end{equation}
and these quantities satisfy the following dispersion relation,
\begin{equation}
    E_{reg}  = \sqrt{\Big(J_{\phi} -  \frac{\sqrt{\lambda}}{2\pi} (b\Delta t)_{reg}\Big)^2 + \frac{\lambda(1 - b^2)}{\pi^2} \cos^2\frac{(\Delta t)_{reg}}{2}} \ .
\end{equation}
The above dispersion relation can be compared with the usual giant magnon dispersion relations, but the interpretation of the above is not clear to us at the moment.

\subsection*{Dispersion relations for pure NS-NS case}

We also try to give a dispersion relation for the pure NS-NS point. In this case the conserved charges become,
\begin{eqnarray}
    E &=& \frac{\sqrt{\lambda}}{\pi} \frac{\sqrt{1 - \mu^2v^2}}{\sqrt{1 - \mu^2}} \int_0^{\infty} \frac{\cosh\rho d\rho}{\sqrt{\cosh^2\rho - k^2\sinh^2\rho}} \ , \nonumber \\ J_{\phi} &=& \frac{\sqrt{\lambda}}{\pi} \frac{(\omega - \mu^2 v)}{\sqrt{(1 - \mu^2)(1 - \mu^2 v^2)}} \int_0^{\infty} \frac{\sinh^2\rho d\rho}{\cosh\rho \sqrt{\cosh^2\rho - k^2\sinh^2\rho}} \ , \nonumber \\ J_{\varphi} &=& \frac{\sqrt{\lambda}}{\pi}\frac{\mu (1 - v^2)}{\sqrt{(1 - \mu^2)(1 - \mu^2 v^2)}} \int_0^{\infty} \frac{\cosh\rho d\rho}{\sqrt{\cosh^2\rho - k^2\sinh^2\rho}}  \ . \nonumber \\
\end{eqnarray}
and the time difference $\Delta t$ is given by,
\begin{eqnarray}
    \Delta t = \frac{2}{\sqrt{(1 - \mu^2)(1 - \mu^2v^2)}} && \Big[ (\omega - \mu^2v) \int_0^{\infty} \frac{\sinh^2\rho d\rho}{\cosh\rho \sqrt{\cosh^2\rho - k^2\sinh^2\rho}} \nonumber \\ && - v(1 - \mu^2) \int_0^{\infty} \frac{\cosh\rho d\rho}{\sqrt{\cosh^2\rho - k^2\sinh^2\rho}}  \ .
\end{eqnarray}
It is clear that in this limit all the quantities are divergent, at least in the IR. Combining $J_{\phi}$ and $\Delta t$, on the other hand, we get,
\begin{equation}
    J_{\phi} - \frac{\sqrt{\lambda}}{2\pi} \Delta t = \frac{\sqrt{\lambda}}{\pi} \frac{ v\sqrt{1 - \mu^2}}{\sqrt{1 - \mu^2v^2}} \int_0^{\infty} \frac{\cosh\rho d\rho}{\sqrt{\cosh^2\rho - k^2\sinh^2\rho}}  \ .
\end{equation}
We find the following regular linear relations among the conserved charges,
\be
E-\mu J_{\varphi} = \frac{1}{v}\Big[ J_{\phi} - \frac{\sqrt{\lambda}}{2\pi} \Delta t \Big].
\ee
This comes as no surprise, since it has been repeatedly noted in literature that the scaling relation among various conserved charegs in the pure NS-NS point becomes linear and considerably simple. One could actually notice that for $\mu=1$ and $v=1$, the above relation can be interpreted as a direct $b=1$ version of (\ref{magnon}).  A similar looking $b=1$ dyonic magnon solution on $\mathbb{R}\times S^3$ was also introduced in \cite{B.Hoare}.
\section{Comments on the soliton solutions}
The solutions discussed in this paper, like usual $AdS$ string solutions, are also `solitonic' in nature. In what follows, we would like to show how these solitons are generated on the worldsheet theory perspective. Let us remind ourselves that the $AdS_3$ space is parameterised by complex two component vectors  $(Z_1, Z_2)$ with,
\be
Z_1= \cosh\rho~ e^{it},~~~Z_2 = \sinh\rho ~e^{i\phi},~~~Z_iZ^{i*} = -1
\ee
This is equivalent to taking a four component real vector
\be
X_i=\left( \cosh\rho\cosh t,\cosh\rho\sin t, \sinh\rho \cos\phi, \sinh\rho\sin\phi  \right),
\ee
so that the Virasoro constraint for the string motion (equivalently the Hamiltonian) can be written in a reduced form,
\be
\dot X_i \dot X^i+X'_i X'^i=-1
\ee
Where the dots and primes are derivatives w.r.t. $\tau$ and $\sigma$. And the combination given by,
\be
\dot X_i \dot X^i-X'_i X'^i=-\cosh^2\rho(\dot t^2 -t'^2)+(\dot\rho^2 -\rho'^2)+\sinh^2\rho(\dot\phi^2-\phi'^2),
\ee
\begin{figure}[!t]
		\centering
		\includegraphics[width=14cm]{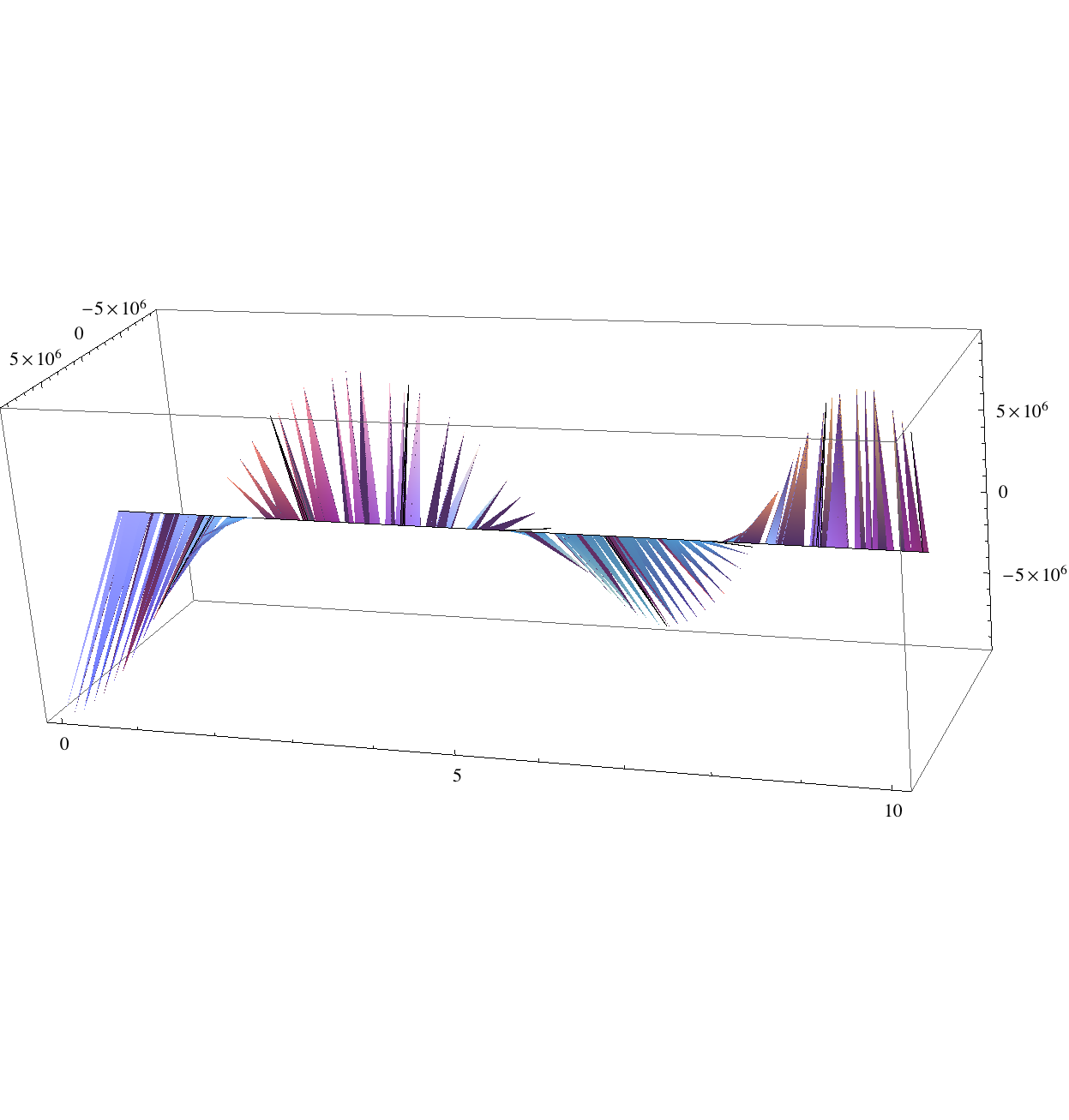}
		\caption{The solution propagating in worldsheet time $\tau = (0,10)$. We have taken $b=0.2, \omega = 0.5, v = 0.7$. The `soliton' structure is quite prominent here.}
\end{figure}

where the solutions for $(\rho, t,\phi)$ involves the information about the NS-NS flux. Now, to see that explicitly, let's start with the $t$ equation of motion (with the choice $c_1=-v$) and use to write,
\be
\frac{\partial g_1}{\partial y} = \frac{(bw-v)}{(1-v^2)}\frac{\hat\gamma^2}{\hat\gamma^2+\sinh^2\hat\beta y},
\ee
which can be integrated to get a solution for the $t$ coordinate in the form,
\be
\cot(t-\tau) = \frac{\hat\gamma}{\sqrt{1-\hat\gamma^2}}\coth\hat\beta y.
\ee
Now, also using the value of $\partial_y g_2 =\frac{\frac{b}{\omega}-v}{(1-v^2)} $, we could write the solution
\be
\phi = \frac{\omega(\tau-v\sigma)+b(\sigma-v\tau)}{(1-v^2)}
\ee
Quite interestingly, one can see that this solution is related to the ones for pure R-R case solution via the worldsheet coordinate transformations,
\be
\sigma \to \omega\sigma + b\tau,~~~\tau \to \omega\tau +b\sigma.
\ee

Using the above, we can write down the solution for the complex $AdS_3$ coordinates,
\bea
Z_1 &=& \pm e^{i\tau}\Big[i\sqrt{1-\hat\gamma^2}-\hat\gamma\coth\hat\beta y   \Big],\nonumber \\
Z_2&=& \pm\frac{\hat\gamma}{\sinh\hat\beta y}~\text{Exp}\Big[i\frac{\omega(\tau-v\sigma)+b(\sigma-v\tau)}{(1-v^2)}\Big].\nonumber \\
\eea
This gives the total string solution embedded on the worldsheet. This provides a generalisation to the solution provided in, for example, \cite{Ryang}.

\section{Summary and conclusion}
In this note, we set out with a humble goal, to analyse the properties and effect of fluxes on a rigidly rotating string in $AdS_3\times S^1$. We motivated this from the non compact sector of $AdS_3\times S^3\times T^4$ with both R-R and NS-NS fluxes. We solved the sigma model equations of motion and classified different string profiles and show how they change with increasing flux value. Everywhere, we have found that the strings react to the flux by `fattening', which has been observed before too. We then find the Noether charges associated to the isometries and find dispersion relations among them via careful regularisation of boundary contributions. These dispersion relations agree with the ones found in the case of $\mathbb{R}\times S^3$ with mixed fluxes, since in both cases, the periodicity in magnon momenta have been spoiled by terms linear in flux parameters. The dispersion relation in pure NS-NS point turns out to be completely linear in charges. We then move to commenting on these solutions in embedding target space coordinates.  

There are a number of modest follow-ups we have in our minds at this point. Since the rigidly rotating strings in $\mathbb{R}\times S^3$ has been discussed carefully, our results provide a `wick rotated' version of the results presented in \cite{B.Hoare}. Surely this can be checked by explicitly wick rotating the gauge fixed sigma model action and solving the equations of motion. Another direct avenue to take would be find out the finite gap solutions for these strings along the lines of $SL(2)$ case \cite{Minahan1}. it would also be interesting to look at these strings from the Neumann-Rosochatius integrable system point of view \cite{HN1}. One of the most intriguing aspect of studies about `mixed flux' backgrounds is that the solutions always simplify at the pure NS-NS point. We also have tried to bring forth such string solutions at $b=1$ point. We hope that the recently uncovered spin-chain picture for the WZW case \cite{purens2} would be able to provide more detailed picture of such cases. Last but not the least, it would be extremely interesting to look at such solutions in a consistent pp-wave limit \cite{Characteristicfrequencies} of backgrounds with NS-NS flux. We hope to focus on these investigations in the near future.

\section*{Acknowledgements} 
Aritra Banerjee (ArB) is supported in part by the Chinese Academy of Sciences (CAS) Hundred-Talent Program, by the Key Research Program of Frontier Sciences, CAS, and by Project 11647601 supported by NSFC. ArB would like to thank CERN Theoretical Physics for kind hospitality during this work was in process. KLP would like to thank Institut {\rm {f\"{u}}r}  Mathematik und Physik, {\rm {Humboldt-Universit\"{a}t} zu Berlin, Germany for hospitality where a part of this work was done.

\end{document}